\documentclass[prl,superscriptaddress,twocolumn,showpacs]{revtex4}

\usepackage{enumerate}
\usepackage{amsfonts,amssymb,amsmath}
\usepackage[]{graphics,graphicx,epsfig}
\usepackage{amsthm}

\usepackage{graphicx}
\usepackage{dcolumn}
\usepackage{natbib}
\usepackage{color}
\usepackage{multirow}

\def\identity{\leavevmode\hbox{\small1\kern-3.8pt\normalsize1}}

\newtheorem{propo}{Proposition}
\newcommand{\be}{\begin{eqnarray}}
\newcommand{\ee}{\end{eqnarray}}
\newcommand{\bpr}{\begin{propo}}
\newcommand{\epr}{\end{propo}}
\newcommand{\bpf}{\begin{proof}}
\newcommand{\epf}{\end{proof}}
\newcommand{\ket}[1]{\left | #1 \right\rangle}
\newcommand{\bra}[1]{\left \langle #1 \right |}

\renewcommand{\epsilon}{\varepsilon}

\begin{document}


\title{Sculpting out quantum correlations with bosonic subtraction}

\author{Marcin Karczewski}   
\affiliation{Faculty of Physics, Adam Mickiewicz University, Umultowska 85, 61-614 Pozna\'n, Poland}

\author{Su-Yong Lee}   
\affiliation{School of Computational Sciences, Korea Institute for Advanced Study, Hoegi-ro 85, Dongdaemun-gu, Seoul 02455, Korea}

\author{Junghee Ryu}   
\affiliation{Centre for Quantum Technologies, National University of Singapore, 3 Science Drive 2, 117543 Singapore, Singapore}

\author{Zakarya Lasmar}   
\affiliation{Faculty of Physics, Adam Mickiewicz University, Umultowska 85, 61-614 Pozna\'n, Poland}

\author{Dagomir Kaszlikowski}   
\affiliation{Centre for Quantum Technologies, National University of Singapore, 3 Science Drive 2, 117543 Singapore, Singapore}

\author{Pawe\l{} Kurzy\'nski}   \email{pawel.kurzynski@amu.edu.pl}   
\affiliation{Faculty of Physics, Adam Mickiewicz University, Umultowska 85, 61-614 Pozna\'n, Poland}
\affiliation{Centre for Quantum Technologies, National University of Singapore, 3 Science Drive 2, 117543 Singapore, Singapore}

\date{\today}


\begin{abstract}
We present a method to extract $M$-partie bosonic correlations from an $N$-partite maximally symmetric state ($M<N$) with the help of successive applications of single-boson subtractions. We also propose an experimental photonic setup to implement it that can be done with the present technologies. 
\end{abstract}

\maketitle


\emph{Introduction.} Recent experimental developments in quantum computing \cite{terhal2018quantum, castelvecchi2017ibm, gomes2018quantum} seem to indicate that universal quantum Turing machine is within our reach. However, the universal quantum computer is not the only way to achieve quantum supremacy. One can envisage useful quantum devices that perform only some specific computational tasks such as boson sampling \cite{aaronson2011computational, zhong201812}, which is a particular example of a wider concept of linear optical quantum computing (LOQC) \cite{knill2001scheme}. LOQC has entered initial stages of technological implementations with integrated optical circuits \cite{crespi2013integrated} offering a much desired miniaturization. It utilizes a purely quantum phenomenon of bosonic indistinguishability and as such is of fundamental significance in our understanding of such systems from information-theoretic point of view.     

In this paper we propose a novel, experimentally feasible technique to extract quantum correlations from maximally symmetric bosonic states within LOQC paradigm. Our technique relies on photonic subtraction that has been already experimentally implemented and shown to be useful for entanglement manipulations in photonic systmes \cite{kim2008scheme,lee2011enhancing, navarrete2012enhancing,parigi2007probing}. Presented results here further fuse developing LOQC technologies with novel ideas about bosonic correlations and their usefulness in quantum computing.

\emph{Basic idea.} Maximally symmetric state represents $N$ bosons, each occupying a different mode $|sym_N\rangle = |1,1,\ldots,1\rangle = a_1^{\dagger}a_2^{\dagger}\ldots a_N^{\dagger}|0\rangle$. Written in the 1st quantization this state is highly correlated: $|sym_N\rangle = \frac{1}{\sqrt{N!}}\sum_{\sigma}|\sigma_{1,2,\ldots, N}\rangle$, where $\sigma$ denotes all possible permutations of particle labels.  An equivalent state of $N$ distinguishable particles does not have any correlations. It has been shown that some of the correlations in $|sym_N\rangle$ can be converted into a usable entanglement in certain scenarios \cite{franco2018indistinguishability, blasiak2018entangling}. We take these ideas further and show that a tailored bosonic subtraction can generate $M$-partite entanglement ($M < N$). 

The intuition behind all the technicalities presented below is that the state $|sym_N\rangle$ contains a plethora of various correlations that can be extracted by removing a number of bosons. It is similar to a sculptor who chisels away a previsualized shape from a piece of marble. Let us see how it works by showing two consecutive chisel strikes. 

The first strike is a subtraction of a boson in a superposition $a'=\sum_{i=1}^{N}\alpha'_i a_i$ after which $|sym_N\rangle$ turns into 
\begin{equation}
\left(\alpha'_1 a_2^{\dagger}\ldots a_N^{\dagger}+\alpha'_2 a_1^{\dagger}a_3^{\dagger}\ldots a_N^{\dagger}+\ldots +\alpha'_N a_1^{\dagger}a_3^{\dagger}\ldots a_{N-1}^{\dagger}\right)|0\rangle.
\end{equation}
The second strike $a''=\sum_{j=1}^{N}\alpha''_j a_j$ further increases the complexity of the resulting state
\begin{equation}
\left(\sum_{i,j=1}^N\alpha'_i\alpha''_j\ldots a_{i-1}^{\dagger}a_{i+1}^{\dagger}\ldots a_{j-1}^{\dagger}a_{j+1}^{\dagger} \ldots\right)|0\rangle.
\end{equation}
Although these strikes are neither hermitian nor unitary they are in fact well established experimentally and they correspond to non-deterministic state manipulations (implementation details are discussed below) \cite{kim2008scheme}. We show next that with a proper choice of $\alpha$'s and repetitive strikes we can sculpt away states with tailored correlations.     

\emph{Bipartite correlations.} Bosonic correlations are subtle because of the intrinsic symmetrization \cite{wiseman2003entanglement}. In quantum information precessing correlations are treated as a resource that can be manipulated and consumed. In this view not every bosonic state is a consumable resource because bosonic wave function is always symmetric. Therefore, a proper quantification of usable bosonic correlations has to be established. There are a few approaches to this problem and we use the one in \cite{Slater1,Slater2,Slater3,Eckert2002}. 

Consider an arbitrary pure bi-bosonic state 
\begin{equation}
|\psi\rangle = \sum_{i,j=1}^d \beta_{ij} a_i^{\dagger}a_j^{\dagger}|0\rangle,
\end{equation}
where $\beta_{ij}$ is a symmetric matrix. It was shown \cite{Slater3} that there is a unitary transformation $a_i^{\dagger}\rightarrow \sum_{i,jk=1}^d\gamma_{ij}c_j^{\dagger}$ bringing $|\psi\rangle$ to $|\tilde{\psi}\rangle$
\begin{equation}
|\tilde{\psi}\rangle = \sum_{i=1}^k \sqrt{\frac{r_i}{2}} c_i^{\dagger 2}|0\rangle.
\end{equation}
In the above formula there are $k\leq d$ positive coefficients $r_i\geq r_{i+1}$ that sum up to one. This is the so-called Slater representation analogous to the ubiquitous Schmidt decomposition \cite{nielsen2002quantum}. The number of non-zero coefficients $k$ is called the Slater rank. The larger the Slater rank the more correlated bosons are and the amount of correlations can be measured by purity defined as $P=\sum_{i=1}^k r_i^2$. Totally uncorrelated states have $P=1$ and the maximally correlated ones come with $P=\frac{1}{d}$.

To warm up let us begin with the state $|sym_4\rangle$ that we will transform into $|\phi_4\rangle$ by consecutive single-boson subtractions
\begin{equation}\label{example}
|sym_4\rangle \rightarrow |\phi_4\rangle=\frac{1}{\sqrt2}(-a_1^\dagger a_2^\dagger + a_3^\dagger a_4^\dagger) \,|0\rangle.
\end{equation}
A simple, single-particle unitary transformation gives its Slater representation 
\begin{equation}
|\tilde{\phi}_4\rangle=\frac{1}{2\sqrt 2}[c_1^\dagger c_1^\dagger+c_2^\dagger c_2^\dagger+c_3^\dagger c_3^\dagger+c_4^\dagger c_4^\dagger]\;|0\rangle
\end{equation}
with purity $P=1/4$ indicating the maximal correlations between two bosons in the four-mode space. The state $|\phi\rangle$ is a result of two chisel strikes $a''a'|sym_4\rangle$ with $a'=\frac{1}{2}(a_1+a_2-a_3-a_4)$ and $a''=\frac{1}{2}(a_1+a_2+a_3+a_4)$. As mentioned before, subtraction is not deterministic and thus $a''a'|sym_4\rangle$ is not normalized (post-selected). If one wants to get any other, non-maximally correlated state this can be done by an appropriate choice of $a',a''$. 

Generalization to $N=2n$ bosons goes as follows. We start with $|sym_{2n}\rangle$ and transform it into the maximally correlated two-boson state in the $2n$ mode space
\begin{eqnarray}
|sym_{2n}\rangle \rightarrow
|\phi_{2n}\rangle = \frac{1}{\sqrt n}\sum_{k=1}^{n}(-1)^ka_{2k-1}^\dagger a_{2k}^\dagger|0\rangle
\end{eqnarray}
via a tailored sequence of subtractions 
\begin{equation}
\prod_{k=1}^{n-1}(a'^{(n-k)} a''^{(n-k)})|sym_{2n}\rangle,
\end{equation}
where
\begin{eqnarray}
&& a'^{(j)}=\frac{1}{2}(a_{2j-1}+a_{2j}+a_{2j+1}+a_{2j+2}) \\
&& a''^{(j)}=\frac{1}{2}(a_{2j-1}+a_{2j}-a_{2j+1}-a_{2j+2}).
\end{eqnarray}
Again, $|\phi_{2n}\rangle$ can be transformed into its Slater form 
\begin{equation}
|\tilde{\phi}_{2n}\rangle=\frac{1}{2\sqrt  n}[c_1^\dagger c_1^\dagger+\ldots+c_n^\dagger c_n^\dagger+c_{n+1}^\dagger c_{n+1}^\dagger+\ldots+c_{2n}^\dagger c_{2n}^\dagger]\;|0\rangle,
\end{equation}
with the Slater rank $P=\frac{1}{2n}$, which is also the minimal possible value in the $2n$-mode scenario (for more details see Appendix).

\emph{Multipartite correlations.} Multipartite bosonic correlations are difficult to analyze because Slater representation may not exist for a given state \cite{Eckert2002}. It is exactly the same situation we encounter for multipartite correlations of distinguishable particles where additionally one has many operationally inequivalent classes of correlations. These difficulties appear in already complicated bosonic correlations.  Therefore, we focus on two non-equivalent classes of qubit correlations: W and GHZ \cite{dur2000three}. To this end, we need to find a proper logical qubits in our multi-bosonic system. For $2n$ modes and $n$ bosons we define the $k$th ($k=1,2,\ldots, n$) logical qubit $\alpha |\mathbf{0}\rangle_k + \beta |\mathbf{1}\rangle_k$ as $(\alpha a_{2k-1}^\dagger+\beta a_{2k}^\dagger)|0\rangle$.

\emph{GHZ correlations.} The goal is $|sym_{2n}\rangle\rightarrow |GHZ_n\rangle$ where  
\begin{eqnarray}
|GHZ_n\rangle &=& \frac{1}{\sqrt2}\left(|\mathbf{00}\ldots\mathbf{0}\rangle+|\mathbf{11}\ldots\mathbf{1}\rangle\right)  \\
&\equiv & \frac{1}{\sqrt2}\left(a_1^\dagger a_3^\dagger \ldots a_{2n-1}^\dagger + a_2^\dagger a_4^\dagger \ldots a_{2n}^\dagger\right)|0\rangle. \nonumber
\end{eqnarray}
The sequence of single-boson subtractions is
\begin{equation}
a^{(1)}a^{(2)} \ldots a^{(n)}|sym_{2n}\rangle,
\end{equation}
where
\begin{eqnarray}
a^{(k)}=\frac{1}{\sqrt{2n}}\left(\sum_{j=1}^n a_{2j-1}  + \sum_{j=1}^n e^{\frac{2\pi i}{n}(j-k)}a_{2j} \right).
\end{eqnarray}

\emph{W correlations.} $W$ state appears to be troublesome as we only managed to find a procedure to obtain it from a $4n$-boson state $|sym_{4n}\rangle$. We have some numerical evidence that this method will not work for $|sym_{2n}\rangle$ but the problem remains open.

The target state is
\begin{eqnarray}
|W_n\rangle &=& \frac{1}{\sqrt{n}}\left(|\mathbf{10}\ldots\mathbf{0}\rangle+|\mathbf{01}\ldots\mathbf{0}\rangle+\ldots +|\mathbf{00}\ldots\mathbf{1}\rangle\right)\nonumber\\
&\equiv &\frac{1}{\sqrt n}(a_2^\dagger a_3^\dagger \ldots a_{2n-1}^\dagger + a_1^\dagger a_4^\dagger \ldots a_{2n-1}^\dagger+ \ldots \nonumber\\
&+&  a_1^\dagger a_3^\dagger \ldots a_{2n}^\dagger)|0\rangle.
\end{eqnarray}
Unlike in all the other cases, this time we need to proceed in two consecutive subtraction steps. The first step is to bring $|sym_{4n}\rangle$ to $|stage_1\rangle$

\begin{eqnarray}
&&|sym_{4n}\rangle\rightarrow |stage_1\rangle = \frac{1}{\sqrt{2^n}}\prod_{i=1}^n\left(\sum_{j=0}^1 a_{2i-j}^\dagger a_{2n+2i-j}^\dagger \right)|0\rangle\nonumber\\
&&
\end{eqnarray}
with the following sequence of subtractions 
\begin{equation}
a^{'(1)}a^{''(1)}\ldots a^{'(n)}a^{''(n)}|sym_{4n}\rangle
\end{equation}
where
\begin{eqnarray}
a^{'(k)}&=&\frac{1}{2}\left(a_{2k-1}+a_{2k}+a_{2n+2k-1}+a_{2n+2k}\right), \\
a^{''(k)}&=&\frac{1}{2}\left(a_{2k-1}-a_{2k}+a_{2n+2k-1}-a_{2n+2k}\right).
\end{eqnarray}

In the final step, $|stage_1\rangle\rightarrow |W_n\rangle$, we subtract exactly $(n-1)$ even-indexed and one odd-indexed particles from the modes with the indices ranging from $2n+1$ to $4n$
\begin{eqnarray}
|W_n\rangle = \mathcal{N}\left(\sum_{i=n+1}^{2n} a_{2i}\right)^{n-1}
\left(\sum_{i=n+1}^{2n} a_{2i-1}\right)|stage_1\rangle,
\end{eqnarray}
where $\mathcal{N}$ is a normalization factor. More detailed description of the above protocols is described in the Appendix.

\begin{figure}[h]
	\centering
	\includegraphics[width=5cm]{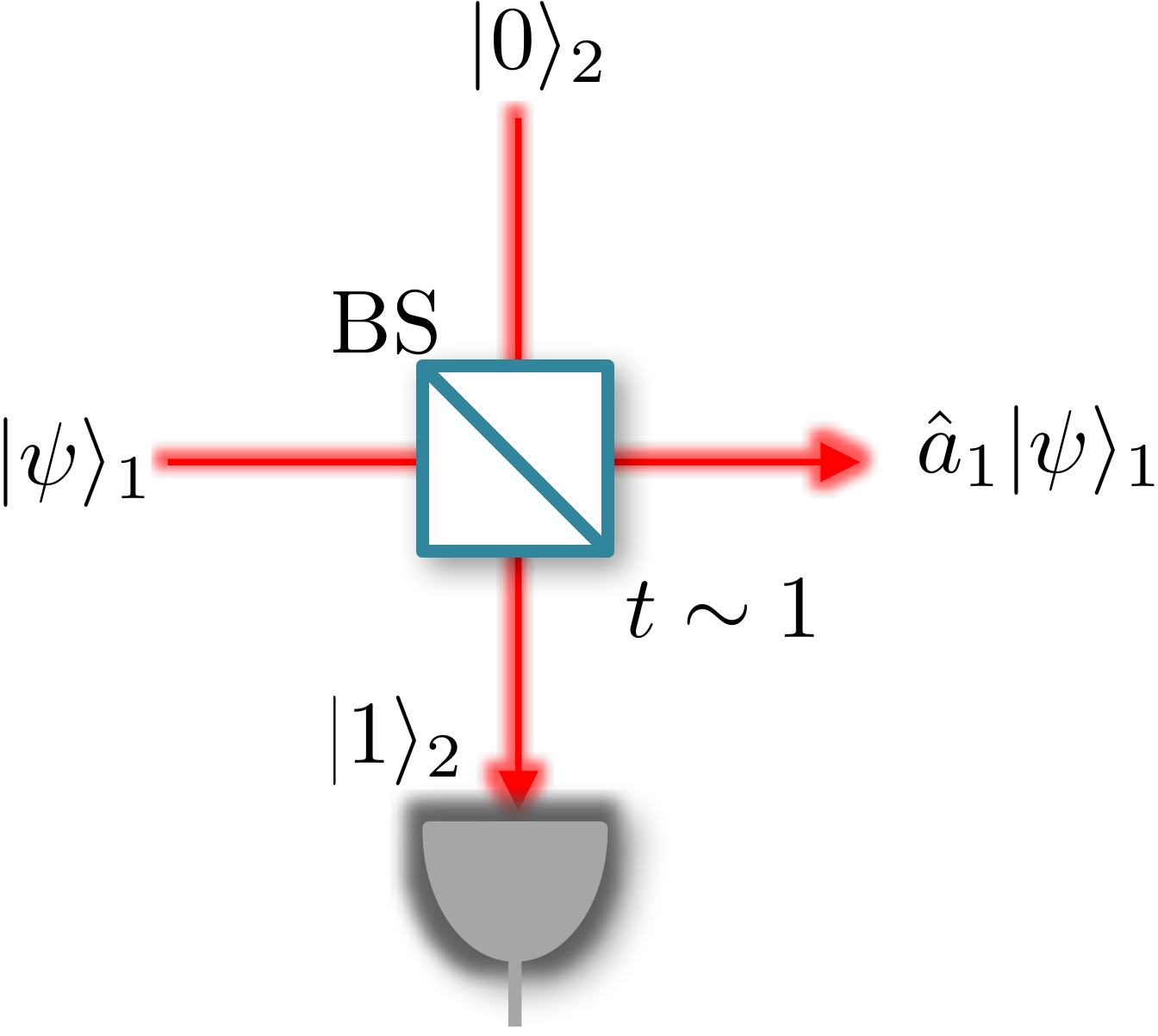}
	\caption{Photonic subtraction: photonic state $|\psi\rangle$ arrives at the input port $1$ of a beamsplitter with high transmittivity ($t\approx 1$). Heralded by a click in the output port $2$, a single photon is subtracted from the state $|\psi\rangle$ and the resulting state $\hat{a}_1|\psi\rangle_1$ comes out in the output port $1$. }
	\label{fig1}
\end{figure}

\emph{Experimental implementation.} The basic operation in our scheme is bosonic subtraction. It was experimentally carried out with photons \cite{kim2008scheme}. The setup uses a simple post-selection procedure (for details see Appendix). A photonic state $|\psi\rangle$ is sent into an input port $1$ of a high transmittivity beamsplitter ($t\approx  1$). A probability that there is a reflection is small and if it happens it is a single photon that triggers a detector in the output port $2$, signaling that the photon was subtracted from the input state $|\psi\rangle$ (see Fig. \ref{fig1}).

Here we discuss our experimental proposal for the sculpting $|sym_4\rangle\rightarrow |\phi_4\rangle$ discussed in Eq. (\ref{example}). 

\begin{figure}[h]
	\centering
	\includegraphics[width=8cm]{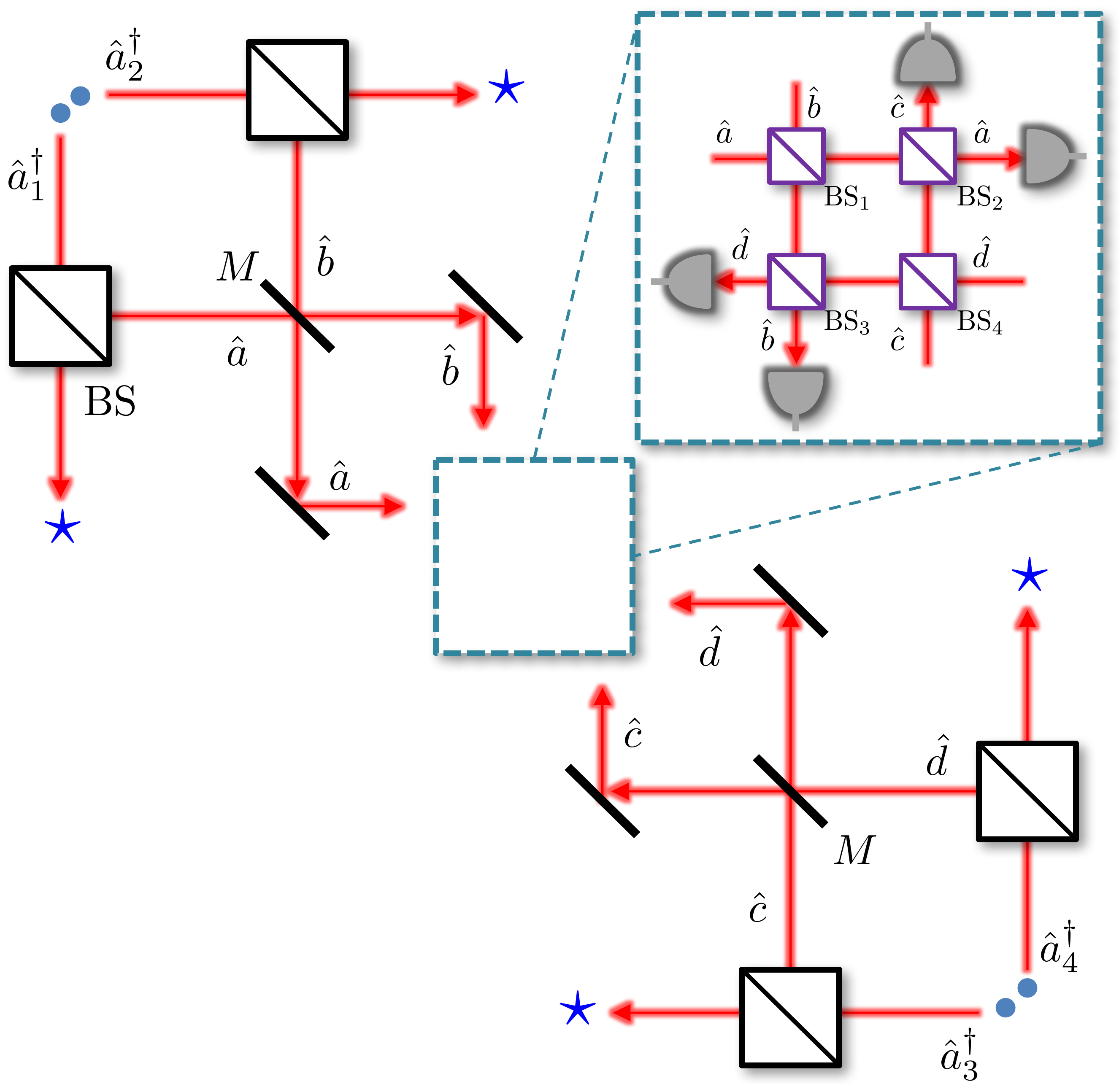}
	\caption{Sculpting $|sym_4\rangle\rightarrow |\phi_4\rangle$: four high transmittivity beamsplitters (BS, black), mirrors (M) and {\it the superposition-subtraction} module in the dashed box are used. The superposition-subtraction module: four $50/50$ beamsplitters (BS, purple) and four detectors are used. If the detector $d$ clicks $|sym_4\rangle$ is post-selected to  $\frac{1}{2}(a+b+c+d)|sym_4\rangle $. The post-selected state  $\frac{1}{2}(a+b-c-d)|sym_4\rangle$ is heralded by the detector $b$.  After a click in $b$, the post-selected state is fed back into the setup thru the input ports $a_1, a_2, a_3, a_4$ and it is further post-selected on the click in $d$, giving the state $\frac{1}{4}(a+b+c+d)(a+b-c-d)|sym_4\rangle$. The opposite sequence of clicks, $d$ followed by $b$, gives the same state.}
	\label{fig3}
\end{figure}

To subtract a photon in a superposition of modes we use the setup sketched in the insert in Fig. \ref{fig3}. It is made of four unbiased beamsplitters [generated by Hamiltonian $-i(a^\dagger b - b^\dagger a)$] and four photon detectors labeled $a,b,c,d$. Photons can enter the interferometer thru four input ports also labeled $a,b,c,d$. If a single photon enters the setup and the detector $b$ clicks then we know that it was in an equal superposition of all four modes $\frac{1}{2}(a^\dagger+b^\dagger-c^\dagger-d^\dagger)|0\rangle$. A detection at $d$ implies that the photon was in $\frac{1}{2}(a^\dagger+b^\dagger+c^\dagger+d^\dagger)|0\rangle$. 


Having mastered subtraction of photons in superposed states we demonstrate how to perform $|sym_4\rangle\rightarrow |\phi_4\rangle$. It is clear that we need to perform a sequence of subtractions of photonic superpositions as described in the preceding paragraph. Such a sequence can be executed with the setup in Fig. \ref{fig3}. We feed it with $|sym_4\rangle$ thru the input ports labeled $a_i^\dagger$ ($i=1,2,3,4$). The beamsplitters (BS) have high transmittivity (like in Fig. \ref{fig1}) and thus the probability that more then a single photon from $|sym_4\rangle$ is subtracted is small. In most of the experimental runs no detectors in the dashed box clicks. However, if the detector $b$ clicks we have the state $\frac{1}{2}(a+b-c-d)|sym_4\rangle$ that we feed back to the setup and wait for $d$'s response, heralding the arrival of the desired state $\frac{1}{4}(a+b+c+d)(a+b-c-d)|sym_4\rangle$ in the output ports labelled by $\star$. For more details on this setup see Appendix. The efficiency of the process depends on the transmittivity $t$ and scales as $\frac{1}{|t|^2}-1$.   

\emph{Conclusions.} Quantum computation and quantum information processing with indistinguishable  particles (bosons, fermions) is an active research area. Recent experimental developments in LOQC and their potential for technological implementations bring us one step closer to commercial quantum computing. Here we consider a generic task of creation and manipulation of quantum correlations contained in bosonic states. 

The main idea of our paper is to extract operationally accessible quantum correlations from a specific $N$-boson state $|sym_N\rangle$. Correlations in this state have been thought to be of no use in quantum protocols until the very recent developments \cite{franco2018indistinguishability, blasiak2018entangling}. Here we show that this state contains a plethora of correlations that can be extracted with a simple operation of particle subtraction. Such an operation is useless for any state made of distinguishable particles or fermions (details in Appendix). 

Our proposal can be experimentally implemented with existing technologies, especially that the basic ingredient is the state $|sym_{N}\rangle$ -- the cornerstone of experimental boson sampling \cite{aaronson2011computational}.

\emph{Acknowledgements.} This work was supported by the National Science Centre in Poland. M.K. acknowledges the Grant No. 2017/27/N/ST2/01858 and the doctoral scholarship no. 2018/28/T/ST2/00148, Z.L. and P.K. acknowledge the Grant No. 2014/14/E/ST2/00585. J.R. and D.K acknowledge the National Research Foundation, Prime Minister’s Office, Singapore and the Ministry of Education, Singapore under the Research Centres of Excellence programme.



%
%
%
%

\appendix
\section{Appendix}
\label{apx:exp}

\section{Bipartite correlations}
\subsection{Four modes}
Consider the state
\begin{equation}
|\Psi_4\rangle=\frac{1}{\sqrt2}(-a_1^\dagger a_2^\dagger + a_3^\dagger a_4^\dagger) \,|0\rangle.
\end{equation}
Notice that it can be expressed as
\begin{equation}
|\Psi_4\rangle=\frac{1}{2\sqrt 2}[c_1^\dagger c_1^\dagger+c_2^\dagger c_2^\dagger-c_3^\dagger c_3^\dagger-c_4^\dagger c_4^\dagger]\;|0\rangle,
\end{equation}
where
\begin{eqnarray}
& &c_1^\dagger=\frac{a_1^\dagger-a_2^\dagger}{\sqrt2},~c_2^\dagger=\frac{a_3^\dagger+a_4^\dagger}{\sqrt2}, \\ & &c_3^\dagger=\frac{a_1^\dagger+a_2^\dagger}{\sqrt2}, ~c_4^\dagger=\frac{a_3^\dagger-a_4^\dagger}{\sqrt2}. \nonumber
\end{eqnarray}
Thus $|\Psi_4\rangle$ is a state of Slater rank 4 with all Slater coefficients of the same absolute value. Its purity equals
\begin{equation}
P(|\Psi_4\rangle)=2(\frac{1}{4})^2+2(-\frac{1}{4})^2=\frac{1}{4},
\end{equation}
which is also the minimal possible value in a four modes scenario.
State $|\Psi_4\rangle$ can be obtained in the following {\it{sculpting}} procedure
\begin{multline}
(a_1+a_2+a_3+a_4)(a_1+a_2-a_3-a_4) \,a_1^\dagger a_2^\dagger a_3^\dagger a_4^\dagger \,|0\rangle=\\
=[a_1^2 +\ldots + a_4^2 + (a_1 a_2 + a_2 a_1) + (-a_1 a_3 + a_3 a_1) + \\
... (-a_3 a_4 - a_4 a_3)] \,a_1^\dagger a_2^\dagger a_3^\dagger a_4^\dagger \,|0\rangle=\\
= 2(-a_1^\dagger a_2^\dagger + a_3^\dagger a_4^\dagger) \,|0\rangle\overset{normalization}{\rightarrow}\frac{1}{\sqrt2}(-a_1^\dagger a_2^\dagger + a_3^\dagger a_4^\dagger) \,|0\rangle.
\end{multline}

\subsection{$2n$ modes ($n\geq2$)}
It is straightforward to generalize the above construction for any even number of modes. To see that, consider the state
\begin{equation}
|\Psi_{2n}\rangle=\frac{1}{\sqrt n}(-a_1^\dagger a_2^\dagger + a_3^\dagger a_4^\dagger+\ldots-a_{2n-3}^\dagger a_{2n-2}^\dagger + a_{2n-1}^\dagger a_{2n}^\dagger) \,|0\rangle.
\end{equation}Since it can be expressed as
\begin{equation}
|\Psi_{2n}\rangle=\frac{1}{2\sqrt  n}[c_1^\dagger c_1^\dagger+\ldots+c_n^\dagger c_n^\dagger-c_{n+1}^\dagger c_{n+1}^\dagger-\ldots-c_{2n}^\dagger c_{2n}^\dagger]\;|0\rangle,
\end{equation}
where
\begin{eqnarray}
& &c_1^\dagger=\frac{a_1^\dagger-a_2^\dagger}{\sqrt2}, \ldots, c_n^\dagger=\frac{a_{2n-1}^\dagger+(-1)^n a_{2n}^\dagger}{\sqrt2},\\ & &c_{n+1}^\dagger=\frac{a_1^\dagger+a_2^\dagger}{\sqrt2}, \ldots, c_{2n}^\dagger=\frac{a_{2n-1}^\dagger-(-1)^n a_{2n}^\dagger}{\sqrt2},  \nonumber
\end{eqnarray}
$|\Psi_{2n}\rangle$ is a state of Schmidt rank $2n$ with all Schmidt coefficients of the same absolute value. Its purity equals
\begin{equation}
P(|\Psi_{2n}\rangle)=n(\frac{1}{2n})^2+n(-\frac{1}{2n})^2=\frac{1}{2n},
\end{equation}
which is the minimal possible value in a $2n$ modes scenario.

State $|\Psi_{2n}\rangle$ can be obtained in the following {\it{sculpting}} procedure
\begin{multline}
(-1)^n(a_1+a_2+a_3+a_4)(a_1+a_2-a_3-a_4)\ldots\\
\ldots(a_{2n-3}+a_{2n-2}+a_{2n-1}+a_{2n})(a_{2n-3}+a_{2n-2}-a_{2n-1}-a_{2n}) \\
a_1^\dagger \ldots a_{2n}^\dagger \,|0\rangle \rightarrow2\,[(-1)^n (a_1 a_2 - a_3 a_4)(a_3 a_4 - a_5 a_6)\ldots \\
(a_{2n-3} a_{2n-2} - a_{2n-1} a_{2n})] \,a_1^\dagger \ldots a_{2n}^\dagger \,|0\rangle=\\
= 2(-a_1^\dagger a_2^\dagger + a_3^\dagger a_4^\dagger+\ldots+-a_{2n-3}^\dagger a_{2n-2}^\dagger + a_{2n-1}^\dagger a_{2n}^\dagger) \,|0\rangle \\ 
\overset{normalization}{\rightarrow}\frac{1}{\sqrt n}(-a_1^\dagger a_2^\dagger + a_3^\dagger a_4^\dagger+\ldots + a_{2n-1}^\dagger a_{2n}^\dagger) \,|0\rangle.
\end{multline}

\section{Multipartite correlations}

Let $a_i^\dagger,\,i\in\{1,2,\ldots,2n\}$ encode n qubits in the dual-rail scheme, i.e. $a_1^\dagger \rightarrow |0\rangle_1,\,a_2^\dagger \rightarrow |1\rangle_1,\,a_3^\dagger \rightarrow |0\rangle_2,\,\ldots,\,a_{2n}^\dagger \rightarrow |1\rangle_n$. We will now discuss how to obtain GHZ an W states by consecutive single particle subtractions applied to the initial state $a_1^\dagger a_2^\dagger\ldots a_{2n}^\dagger|0\rangle$.

\subsection{Generation of the GHZ state}
Before we proceed to the general case, let us take a look at a specific example: tripartite GHZ state. The idea behind its generation relies on the properties of the third root of unity, denoted $\omega=e^{\frac{2\pi i}{n}}$. One can check that
\begin{multline}
(a_1+a_3+a_5+a_2+\omega a_4+ \omega ^2 a_6)(a_1+a_3+a_5+\omega^2 a_2+ a_4+ \omega  a_6)\\
(a_1+a_3+a_5+\omega a_2+\omega^2 a_4+ a_6) a_1^\dagger a_2^\dagger\ldots a_{6}^\dagger|0\rangle=\\
=[ 6\; a_1 a_3 a_5 +2(1+\omega+\omega^2)(a_1 a_2 a_3 + a_1 a_2 a_5 + \ldots a_3 a_5 a_6)+\\
+2(1+\omega+\omega^2)(a_1 a_2 a_4 + a_1 a_2 a_6 + \ldots a_4 a_5 a_6)+6\;\omega^3a_2 a_4 a_6 ] \\ a_1^\dagger a_2^\dagger\ldots a_{6}^\dagger|0\rangle=6\;[ a_1 a_3 a_5 + a_2 a_4 a_6 ]\;  a_1^\dagger a_2^\dagger\ldots a_{6}^\dagger|0\rangle \\
\overset{normalization}{\rightarrow}\frac{1}{\sqrt2}(a_2^\dagger a_4^\dagger a_6^\dagger+ a_1^\dagger a_3^\dagger a_5^\dagger )|0\rangle
\end{multline}

In case of the n-partite GHZ state we will show that the state
\begin{equation}
|GHZ_n\rangle=\frac{1}{\sqrt2}(a_1^\dagger a_3^\dagger \ldots a_{2n-1}^\dagger +(-1)^{n+1} a_2^\dagger a_4^\dagger \ldots a_{2n}^\dagger)|0\rangle.
\end{equation}
can be prepared by performing the following single-particle subtractions:
 \begin{multline}
(a_1+a_3+\ldots a_{2n-1}+a_2 +e^{\frac{2\pi i}{n}}a_4 +e^{2\frac{2\pi i}{n}}a_6 +\ldots e^{(n-1)\frac{2\pi i}{n}} a_{2n}) \\
(a_1+a_3+\ldots a_{2n-1}+e^{(n-1)\frac{2\pi i}{n}} a_2 + a_4 +e^{\frac{2\pi i}{n}}a_6 +\ldots e^{(n-2)\frac{2\pi i}{n}} a_{2n}) \\
\vdots\\
(a_1+a_3+\ldots a_{2n-1}+e^{\frac{2\pi i}{n}}a_2 +e^{3\frac{2\pi i}{n}}a_4 +\ldots  a_{2n}) a_1^\dagger a_2^\dagger\ldots a_{2n}^\dagger|0\rangle\\
\end{multline}
It is easy to see that the term $a_2^\dagger a_4^\dagger \ldots a_{2n}^\dagger$ remains after the subtractions. This is also true for $a_1^\dagger a_3^\dagger \ldots a_{2n-1}^\dagger$ since 
\begin{equation}
e^{\frac{2\pi i}{n}0}e^{\frac{2\pi i}{n}1}\ldots e^{\frac{2\pi i}{n}(n-1)}=(-1)^{n+1}.
\end{equation}
What remains to be shown is that all the other terms vanish after the subtractions. To see this, consider the term that emerges when we take $a_{i_1}$ with coefficient $c_{i_1}$ from the first subtraction,  $a_{i_2}$ with coefficient $c_{i_2}$ from the second and so on.  Let exactly k of these annihilation operators have even indices, denoted $e_1, ..., e_k$. Then notice that you could have taken each even-indexed annihilation operator $a_{e_j}$ with coefficient $ e^{\frac{2\pi i}{n}} c_{e_j}$ instead of $c_{e_j}$ (this means that we just take them from different subtraction; it is allowed because the relative phases between the coefficients are fixed). Thus the sum over all products of coefficients that lead to the subtraction of $a_{i1}...a_{in}$, which we denote as $S$, must satisfy
\begin{equation}
S= e^{\frac{2\pi i}{n} k} S
\end{equation} 
Unless $ e^{\frac{2\pi i}{n} k}=1$ (so unless k=0 or k=n), this means that $S = 0$, so that the corresponding term vanishes, which completes the proof.

\subsection{Generation of the W state}

In case of the generation of the W state 
\begin{eqnarray}
|W_n\rangle &=& \frac{1}{\sqrt n}(a_2^\dagger a_3^\dagger \ldots a_{2n-1}^\dagger + a_1^\dagger a_4^\dagger \ldots a_{2n-1}^\dagger+ \ldots + \nonumber \\ 
& & a_1^\dagger a_3^\dagger \ldots a_{2n}^\dagger)|0\rangle 
\end{eqnarray}
we have not been able to find a procedure that relied on single-particle subtractions from $a_1^\dagger a_2^\dagger\ldots a_{2n}^\dagger|0\rangle$. However, it could be done if we start with twice the number of particles, i.e. initial state  $a_1^\dagger a_2^\dagger\ldots a_{4n}^\dagger|0\rangle$. The procedure is twofold:\\
Step 1. Create $n$ qubits using particles indexed from 1 to $2n$ and their copy with particles indexed from $2n+1$ to $4n$.\\
This can be done  by the following subtractions:
 \begin{multline}
 (a_1+a_2+a_{2n+1}+a_{2n+2})(a_1-a_2+a_{2n+1}-a_{2n+2}) \\
 (a_3+a_4+a_{2n+3}+a_{2n+4})(a_3-a_4+a_{2n+3}-a_{2n+4})\\
 \vdots\\
 (a_{2n-1}+a_{2n}+a_{4n-1}+a_{4n})(a_{2n-1}-a_{2n}+a_{4n-1}-a_{4n}) \\ a_1^\dagger a_2^\dagger\ldots a_{4n}^\dagger|0\rangle \overset{normalization}{\rightarrow}\frac{1}{\sqrt{2^n}}(a_1^\dagger a_{2n+1}^\dagger + a_2^\dagger a_{2n+2}^\dagger) \\
 (a_3^\dagger a_{2n+3}^\dagger + a_4^\dagger a_{2n+4}^\dagger)\ldots (a_{2n-1}^\dagger a_{4n-1}^\dagger + a_{2n}^\dagger a_{4n}^\dagger)|0\rangle
 \end{multline}
 Example: threepartite W state
 \begin{multline}
  (a_1+a_2+a_{7}+a_{8})(a_1-a_2+a_{7}-a_{8})
 (a_3+a_4+a_{9}+a_{10}) \\ (a_3-a_4+a_{9}-a_{10})
(a_5+a_6+a_{11}+a_{12})(a_5-a_6+a_{11}-a_{12}) \\
a_1^\dagger a_2^\dagger\ldots a_{12}^\dagger|0\rangle=8(a_1 a_7 + a_2 a_8)(a_3 a_{9} + a_4 a_{10}) \\
(a_5 a_{11} + a_6 a_{12})a_1^\dagger a_2^\dagger\ldots a_{12}^\dagger|0\rangle \overset{normalization}{\rightarrow} \\ 
\frac{1}{\sqrt{8}}(a_1^\dagger a_7^\dagger + a_2^\dagger a_8^\dagger)(a_3^\dagger a_{9}^\dagger + a_4^\dagger a_{10}^\dagger)(a_5^\dagger a_{11}^\dagger + a_6^\dagger a_{12}^\dagger)|0\rangle
\end{multline}
Step 2. Subtract exactly $(n-1)$ even-indexed and 1 odd-indexed particles from the ``copy'' part of the state obtained in step 1
\begin{multline}
(a_{2n+2}+a_{2n+4}+\ldots+a_{4n})^{n-1}(a_{2n+1}+a_{2n+3}+\ldots+a_{4n-1}) \\
\frac{1}{\sqrt{2^n}}(a_1^\dagger a_{2n+1}^\dagger + a_2^\dagger a_{2n+2}^\dagger)(a_3^\dagger a_{2n+3}^\dagger + a_4^\dagger a_{2n+4}^\dagger)\ldots \\
(a_{2n-1}^\dagger a_{4n-1}^\dagger + a_{2n}^\dagger a_{4n}^\dagger)|0\rangle \overset{normalization}{\rightarrow} \\
\frac{1}{\sqrt n}(a_2^\dagger a_3^\dagger \ldots a_{2n-1}^\dagger + a_1^\dagger a_4^\dagger \ldots a_{2n-1}^\dagger+ \ldots +  a_1^\dagger a_3^\dagger \ldots a_{2n}^\dagger)|0\rangle
\end{multline}
Example: threepartite W state (continued)
\begin{multline}
(a_{8}+a_{10}+a_{12})^{2}(a_{7}+a_{9}+a_{11})
[\frac{1}{\sqrt{8}}(a_1^\dagger a_7^\dagger + a_2^\dagger a_8^\dagger) \\
(a_3^\dagger a_{9}^\dagger + a_4^\dagger a_{10}^\dagger)(a_5^\dagger a_{11}^\dagger + a_6^\dagger a_{12}^\dagger)]|0\rangle\\
\overset{normalization}{\rightarrow} \frac{1}{\sqrt 3}(a_2^\dagger a_3^\dagger  a_{5}^\dagger + a_1^\dagger a_4^\dagger  a_{5}^\dagger+   a_1^\dagger a_3^\dagger a_{5}^\dagger)|0\rangle
\end{multline}
Notice that one can easily tweak the second step of the above procedure to create Dicke states, i.e. symmetric states of a given number of 0 and 1 bits.

\section{Description of subtraction operation}
The beamsplitter in Fig.~\ref{fig1} in the main text is formulated as $\hat{B}_{12} (\theta, \phi) = \exp \left[ \frac{\theta}{2} \left( \hat{a}_{1}^{\dagger} \hat{a}_{2} e^{i \phi} - \hat{a}_{1} \hat{a}_{2}^{\dagger} e^{-i \phi} \right)\right]$ with $r=\sin \left(\theta/2 \right) e^{i \phi}$ and $t=\cos\left(\theta/2 \right)$. For convenience, we fix the phase difference $\phi=0$. With high transmittivity $t \approx 1$ (or equivalently a parameter $\theta$ is small), a single photon subtraction is implemented if a single photon is detected at the output port $2$:
\begin{eqnarray}
_{2}\bra{1} \hat{B}_{12} (\theta, 0) \ket{\psi}_{1} \ket{0}_{2} &\approx& {}_{2}\bra{1} \left( \hat{\openone}- \frac{r^*}{t} \hat{a}_{1} \hat{a}^{\dagger}_{2} \right) \ket{\psi}_{1} \ket{0}_{2} \nonumber \\
&\approx& \hat{a}_{1} \ket{\psi}_{1}.
\end{eqnarray}

\section{Analysis of Experimental scheme}
We here show in detail how to implement the following sculpting procedure depicted in Fig.~\ref{fig3}: $\left( \hat{a}_{1} + \hat{a}_{2} + \hat{a}_{3} + \hat{a}_{4} \right) \left( \hat{a}_{1} + \hat{a}_{2} - \hat{a}_{3} - \hat{a}_{4}  \right) \ket{sym_4}$.

Let us consider a conditional detection part (see the insert in Fig.~\ref{fig3}). With high transmittivity beamsplitters, the input state of the detection part can be put as
\begin{eqnarray}
&&( \hat{\openone} + \hat{a}^{\dagger}\hat{a}_1 ) ( \hat{\openone} + \hat{b}^{\dagger}\hat{a}_2 ) ( \hat{\openone} + \hat{c}^{\dagger}\hat{a}_3 ) ( \hat{\openone} + \hat{d}^{\dagger}\hat{a}_4 ) \nonumber \\
&& \times \ket{sym_4}_{a_1 a_2 a_3 a_4} \ket{0}_{a b c d},
\label{eq:apx_inputstate}
\end{eqnarray}
where $a_j$ for $j=1,2,3,4$ denotes optical modes for input state and the modes $a,b,c,d$ describe the detection parts. Four symmetric beamsplitters ($50/50$) are used in the detection part in the insert in Fig.~\ref{fig3}, each of which is operated as
\begin{eqnarray*}
&\text{BS}_1&: 
\begin{cases}
\hat{a}^{\dagger} \rightarrow \frac{1}{\sqrt{2}} \left( \hat{a}^{\dagger} + \hat{b}^{\dagger} \right) \\
\hat{b}^{\dagger} \rightarrow \frac{1}{\sqrt{2}} \left( \hat{b}^{\dagger} - \hat{a}^{\dagger} \right)
\end{cases}
\text{BS}_2: 
\begin{cases}
\hat{a}^{\dagger} \rightarrow \frac{1}{\sqrt{2}} \left( \hat{a}^{\dagger} - \hat{c}^{\dagger} \right) \\
\hat{c}^{\dagger} \rightarrow \frac{1}{\sqrt{2}} \left( \hat{c}^{\dagger} + \hat{a}^{\dagger} \right)
\end{cases}
\nonumber \\
&\text{BS}_3&: 
\begin{cases}
\hat{b}^{\dagger} \rightarrow \frac{1}{\sqrt{2}} \left( \hat{b}^{\dagger} + \hat{d}^{\dagger} \right) \\
\hat{d}^{\dagger} \rightarrow \frac{1}{\sqrt{2}} \left( \hat{d}^{\dagger} - \hat{b}^{\dagger} \right)
\end{cases}
\text{BS}_4: 
\begin{cases}
\hat{c}^{\dagger} \rightarrow \frac{1}{\sqrt{2}} \left( \hat{c}^{\dagger} + \hat{d}^{\dagger} \right) \\
\hat{d}^{\dagger} \rightarrow \frac{1}{\sqrt{2}} \left( \hat{d}^{\dagger} - \hat{c}^{\dagger} \right).
\end{cases}
\end{eqnarray*}
After the above beamsplitters operations, the input state (\ref{eq:apx_inputstate}) becomes
\begin{eqnarray}
&&\left[ \hat{\openone} + \frac{1}{2} \left\{ \left( \hat{a}^{\dagger} - \hat{c}^{\dagger} \right) + \left( \hat{b}^{\dagger} + \hat{d}^{\dagger} \right) \right\} \hat{a}_1 \right] \nonumber \\
&&\times \left[ \hat{\openone} + \frac{1}{2} \left\{ \left( \hat{b}^{\dagger} + \hat{d}^{\dagger} \right) - \left( \hat{a}^{\dagger} - \hat{c}^{\dagger} \right) \right\} \hat{a}_2 \right] \nonumber \\
&& \times \left[ \hat{\openone} + \frac{1}{2} \left\{ \left( \hat{c}^{\dagger} + \hat{a}^{\dagger} \right) + \left( \hat{d}^{\dagger} - \hat{b}^{\dagger} \right) \right\} \hat{a}_3 \right] \nonumber \\
&&\times \left[ \hat{\openone} + \frac{1}{2} \left\{ \left( \hat{d}^{\dagger} - \hat{b}^{\dagger} \right) - \left( \hat{c}^{\dagger} + \hat{a}^{\dagger} \right) \right\} \hat{a}_4 \right] \nonumber \\
&&\times \ket{sym_4}_{a_1 a_2 a_3 a_4} \ket{0}_{a b c d}.
\label{eq:apx_output}
\end{eqnarray}

Let us consider that a single photon is detected in $d$ output mode. As a result, the operators involving $\hat{d}^{\dagger}$ term of (\ref{eq:apx_output}) only survive, and the input state is post-selected as
\begin{eqnarray}
&&{}_{abc}\bra{0} {}_{d}\bra{1} \hat{\mathcal{O}}_{\text{BS}} \ket{sym_4}_{a_1 a_2 a_3 a_4} \ket{0}_{abcd} \nonumber \\
&&= \frac{1}{2} \left( \hat{a}_1 + \hat{a}_2 + \hat{a}_3 + \hat{a}_4 \right) \ket{sym_4}_{a_1 a_2 a_3 a_4},
\label{eq:apx_dclick}
\end{eqnarray}
where $\hat{\mathcal{O}}_{\text{BS}}$ denotes the operators part in (\ref{eq:apx_output}). Similarly, for the click of the mode $b$, the post-selected states reads
\begin{eqnarray}
&&{}_{acd}\bra{0} {}_{b}\bra{1} \hat{\mathcal{O}}_{\text{BS}} \ket{\psi}_{a_1 a_2 a_3 a_4} \ket{0}_{abcd} \nonumber \\
&&= \frac{1}{2} \left( \hat{a}_1 + \hat{a}_2 - \hat{a}_3 - \hat{a}_4 \right) \ket{\psi}_{a_1 a_2 a_3 a_4}.
\label{eq:apx_bclick}
\end{eqnarray}
Therefore, as we described in the main text, if the detector of mode $b$ clicks, then the input state is post-selected to $\frac{1}{2}\left( \hat{a}_1 + \hat{a}_2 - \hat{a}_3 - \hat{a}_4 \right) \ket{sym_4}_{a_1 a_2 a_3 a_4}$. Next, this state is fed back into the input port, and is post-selected again by the click of the detector of mode $d$, which results in the final state $\frac{1}{4}\left( \hat{a}_1 + \hat{a}_2 + \hat{a}_3 + \hat{a}_4 \right) \left( \hat{a}_1 + \hat{a}_2 - \hat{a}_3 - \hat{a}_4 \right) \ket{sym_4}_{a_1 a_2 a_3 a_4}$.



\section{Subtraction of fermions}

Here we show that the sculpting protocol does not work for fermions. Consider an anti-symmetric state of $N$ fermions
\begin{equation}
|asym_N\rangle = f^{\dagger}_1\ldots f^{\dagger}_N|0\rangle,
\end{equation}
where $f^{\dagger}_i$ are fermionic creation operators for modes $i=1,\ldots,N$. Next, consider a fermionic subtraction $\tilde{f}|asym_N\rangle$. The operator $\tilde{f}$ can be always written as
\begin{equation}\label{fsub}
\tilde{f} = \alpha\tilde{f}_{sup} + \beta\tilde{f}_{\perp}. 
\end{equation}
The operator $\tilde{f}_{sup}$ is supported on the same modes as the state $|asym_N\rangle$, i.e., 
\begin{equation}
\tilde{f}_{sup} = \sum_{i = 1}^{N} \gamma_{i} f_{i},
\end{equation}
whereas $\tilde{f}_{\perp}$ is supported on some other, orthogonal modes. This means that $\tilde{f}_{\perp}|asym_N\rangle=0$ and $\tilde{f}|asym_N\rangle=\alpha \tilde{f}_{sup}|asym_N\rangle$. However, due to the fact that the state $|asym_N\rangle$ is invariant under single-particle transformations $Uf_i = \tilde{f}_i$, it can be written as 
\begin{equation}
|asym_N\rangle = \tilde{f}^{\dagger}_{sup}\tilde{f}^{\dagger}_{2}\ldots \tilde{f}^{\dagger}_{N}|0\rangle,
\end{equation}
where the operators $\tilde{f}^{\dagger}_i$ ($i=2,\ldots,N$) are mutually orthogonal and orthogonal to $\tilde{f}^{\dagger}_{sup}$. As a result 
\begin{equation}
\tilde{f}|asym_N\rangle=\alpha \tilde{f}^{\dagger}_{2}\ldots \tilde{f}^{\dagger}_{N}|0\rangle \equiv |asym_{N-1}\rangle.
\end{equation}
The state $|asym_{N-1}\rangle$ is uncorrelated, because it is described by a single Slater determinant \cite{Eckert2002}. Finally, note that a sequence of $M$ fermionic subtractions reduces the state $|asym_N\rangle$ to $|asym_{N-M}\rangle$.

\end{document}